# REVIEW OF MONITORING TOOLS FOR E-LEARNING PLATFORMS

Ali Alowayr and Atta Badii

Department of Computer Science, Reading University, Reading City, UK

## ABSTRACT

*The advancement of e-learning technologies has made it viable for developments in education and technology to be combined in order to fulfil educational needs worldwide. E-learning consists of informal learning approaches and emerging technologies to support the delivery of learning skills, materials, collaboration and knowledge sharing. E-learning is a holistic approach that covers a wide range of courses, technologies and infrastructures to provide an effective learning environment. The Learning Management System (LMS) is the core of the entire e-learning process along with technology, content, and services. This paper investigates the role of model-driven personalisation support modalities in providing enhanced levels of learning and trusted assimilation in an e-learning delivery context. We present an analysis of the impact of an integrated learning path that an e-learning system may employ to track activities and evaluate the performance of learners.*

## 1. INTRODUCTION

In recent times, there have been advances in the development of social software technology in particular in the field of education [1]. Blended learning strategies can optimise the integration of multi-modal, multi-channel and multi-source learning which includes online and traditional learning; this helps learners develop and improve their learning autonomy and to self-manage to best suit their learning style, lifestyle and work style. Such software applications are generally developed on web 2.0 tools, for example m-learning applications, twitter, YouTube, slide share, Picasa, media wiki, etc. In the field of education this software is used to help teachers to monitor students' activities. In specific terms, e-learning is based not only on distributed learning, online learning, virtual learning, web based or networked learning but also on testing and evaluating the best feedback, intervention and the interaction of some platforms in e-learning environments between the instructor and learner [2].

Interest in e-learning is on an upward trend particularly for those already in full-time employment and keen to continue their education and/or professional training. The percentage of companies planning to provide e-learning support for their staff has risen from 38.5% in 2007 to 51% in 2011 [3]. Whether the mode of delivery would be online or offline, synchronous or asynchronous via standalone or networked computers or other electronic devices, learning would be delivered using electronic devices [4]. Teacher-student interaction with the help of electronic media and application tools is termed as the concept of e-learning [1]. Going ahead in the line of e-learning modalities, it is important to discuss web-based learning environments and investigate the trade-off between factors, e.g. cognitive and learning styles and teaching delivery models that can influence learning outcomes.

The Learning Management System (LMS) does not only provide the facility to the instructors for assessment and the measurement of the performance of the students in an interactive learning process [5]. This paper is willing to review Monitoring and Analysis Tool for the E-learning Program (MATEP) which is used in this area of e-learning. The paper is also underpinned with an empirical analysis of the concept along with this tool by conducting experiments with student groups to investigate how best one can optimise learning pathways e.g. by creating an environment of real-time interaction between learners and their instructors. A comparative analysis of the different platforms is performed. The results of those experiments are discussed with the analysis of the findings, recommendations and conclusions thereon.

## 2. Literature Review

### 2.1. E-Learning background

E-learning networked information and communications technology is a modern process in the teaching and learning of undergraduate students. There are a number of terms that are used for the purpose of describing this online mode of teaching and learning. The method most widely uses online activities, virtual learning, and a web based learning network and distributed teaching [6]. In this type of educational delivery, information and communications technology are followed in synchronous and asynchronous modes. If a close scrutiny of the entire process is made, it is clear that these communication technologies different educational processes and as such cannot be used simultaneously [7]. It is clear from the information available, the role communication and information technology to provide opportunities in the areas of storing, capturing and disseminating information [6]. Whether it is able to a large variety of information and its quality on learning, the nature and the degree influence etc. is subject to debate and discussion. With the event of media tools and equipment, radio and television are areas that could be used for debate [7]. Video clips used for role-play based learning have made the learning and teaching experience more appealing and revolutionary [6]. It has been seen that students who generally want to stay away from school are highly motivated by innovative teaching methods. Moving images and educational instructions given through multi media have aided in increasing the ability of instructors to represent information, in many ways within the school [7]. With the popularity of radio and television and such other media, it is assumed that every media type has the capability and potential to increase, direct, and shape individuals' abilities to communicate and instruct.
Computers are second in line to provide technological innovation, both with the help of the latest software and hardware, in the area of an effective e-learning process. Though, computers have been used for educational purposes since the 1960s and 70s they were taken up more enthusiastically among e-learning volunteers, students and instructors during 2000 with the help of the Internet [8]. However, there are many researchers who are of the opinion that the impact of computer technology on learning is less than its economic benefits and in some cases they did not show any benefits on learning. Most of the researchers are of the opinion that learning is acquired not with improvement in technology alone but with the content presented in the medium.

### 2.2. Tools for Monitoring and Analysis

The concept of e-learning with the use of computer related tools that run on the Web 2.0 framework will be discussed in this literature review in order to better understand them. Instructors and learners throughout the world who have used e-learning have made this mode a crucial part to fulfil their educational needs [9]. The teachers' only concern is that they are not able to track the activities of learners in terms of what they are learning with the e-learning process. Similarly, the learners are having problems related to course content and its effectiveness when they take part in the e-learning process [10]. Learners' require specific knowledge about the content that has been taught along with assurance of its accuracy and authenticity. Consequently, the Learners' management system that relies on web servers is developed using application tools and other resources such as a Monitoring and Analysis Tool for the E-learning Program (MATEP) to resolve both instructors' and learners' concerns [11]. MATEP tools run over the Web servers with log files where every activity of learners can be captured, and is designed to help instructors to track their learners' activities online [12]. MATEP tools are enriched with the necessary information sourced from academic and social demographic areas. MATEP is a web application that provides a platform for the LMS and e-learning could be performed independently. It can provide information that is up dated automatically at a frequency desired by the administration.

### 2.3. Other Frameworks

Another framework for the delivery of e-learning process could be the virtual environment which is closely linked to the e-learning process. Avatars are the system that works online for representations of the self and virtual world and that are designed to increase interaction between instructors and students. The users of avatars are able to create a personality that is visible to the virtual world that allows the users to use imaginary experiences in order to transform the actual world in which they live. Avatars in the virtual environment are based on the social networking and entertainment purposes now well established.

The use of educational contexts in the Avatar system is in its early stage of development process along with the applications [13]. Some of the applications of such a system of the Avatar in a virtual framework are early simulation games, activities, such as the Sim series and other specialist uses mainly in science and mathematics. There are researchers who are of the opinion that Avatars can be used for virtual discussion in online object-oriented environments [14 -17], claiming that such environments have the potential to enhance social interaction and support connectedness, particularly among those individuals who may have difficulty communicating on a face-to-face basis.

## 3. Experiment

The experiment for this study will be in the form of primary research that clearly explains the effectiveness of the e-learning program [13]. The tools and framework used to perform e-learning will be evaluated on the criterion of the personal experience of instructors and learners. The teachers' concern that they are not able to trace the learners' activity while their entire teaching session and the learners' question about the validity and reliability of the content delivered in the e-learning process [18]. The experiment will make use of samples of a few instructors and e-learners who are using MATEP and Avatars as their virtual framework for learning [19].

A questionnaire will be developed to be completed according to the responses made by the sample user group. The sample size is set to be 150 people consisting of both instructors and learners, both male and female gender of different ages. The questions will be based on the pattern of direct asking and will not use anything that would be disguised to reveal any relevant matter [20]. There will be five open-ended questions which means the reply to such questions will only be either 'Yes' or 'No' or direct response to the question. Questions will be the same for all the respondents and the language will be English. Responses received from the questions will be analysed using pie charts, bar diagrams, etc. and the discussion on each response will be made. The questions put to the respondents are as follows:

### 3.1. Questionnaire

Q1. What e-learning techniques do you follow?

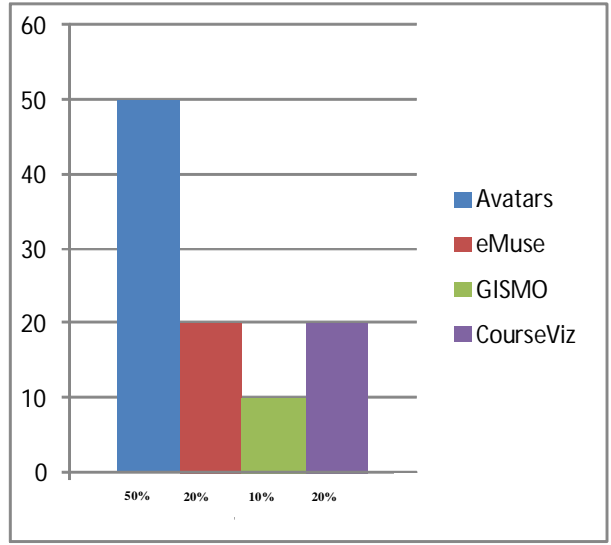

Figure 1. Techniques of e-Learning

Discussion: This is a question asked both to the instructors and learners who are following different techniques in the e-learning process. As the graph reveals, from the four categories of techniques: Avatars, eMUSE, Gismo, CourseViz, etc Avatars are the mostly commonly used technique. This virtual environment technique that is basically used for more dedicated virtual interactions is popular among other techniques with 50% of the respondents using it. Other techniques eMUSE and Courseviz both had a 20% positive response; GISMO had the least of all i.e. only 10% responses.

Q2. What technology seems to be best for the delivery of an effective e-learning program?

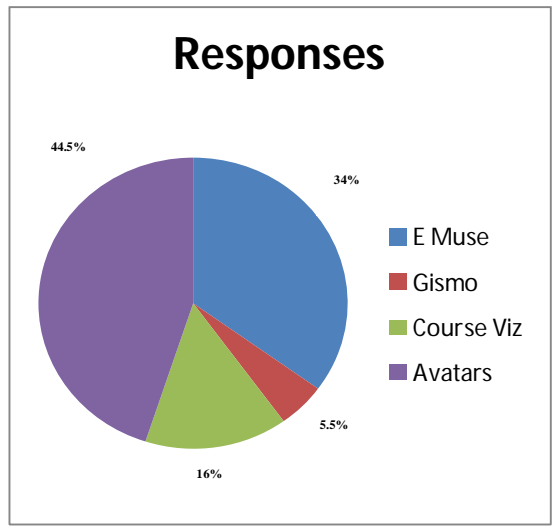

Figure 2. Preferred techniques of e-Learning

Discussion: As shown in the figure 2, Avatars techniques prove to be the best and this explains the reason for its popularity among the users worldwide. The pie graph above shows that about 44.5% of the respondents have given positive response in favour of Avatars. Avatars are popular since they are user friendly and the instructor, when interacting with a learner, could also track their activity in the simulated environment. EMUSE is next in the order of preference which is about 34%, and then the others, it also means that eMUSE is better when using it for sharing knowledge online.

Q3. What are the best virtues you have found in following the MATEP (performance measurement) or other such framework?

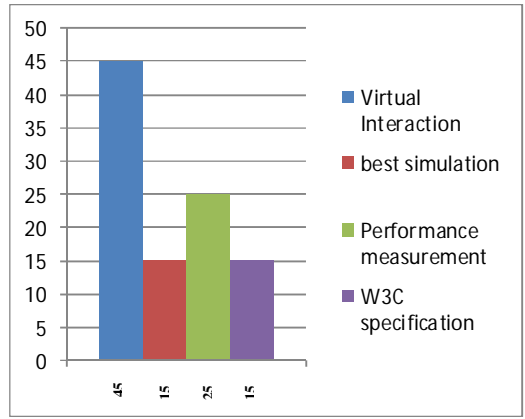

Figure 3. LMS platform comparison

Discussion: The best thing that is provided with the measuring and analysis tools for online program is the creation and management of virtual online environment that exactly simulates to a learning environment. Performance measurement (MATEP) is the most significant concern of the people who are attached in the process of e-learning. Instructors and professors with the help of MATEP tools become able to evaluate learner performances and this is the reason, that criterion for performance measurement is second in the line after virtual interaction.

Q4. What are problems you are still facing while following Web 2.0 frame works?

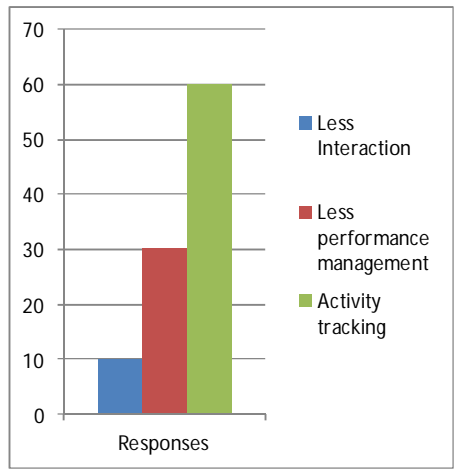

Figure 4. Problems from E-Learning research methods.

Discussion: There are several problems and difficulties still left among the users of virtual techniques who provide online teaching and learning guidance. MATEP tools have proven to be an efficient tool in solving e-learning programs but as the requirements of the e-learning frameworks are increasing becoming difficult for people to follow efficiently. There are results showing that activity tracking (MATEP) is still a problem even after using different tools such as Avatars, eMUSE, etc. It is problematic to track exactly the activities of the learners. More than 50% of the respondents blamed the present MATEP tools for not delivering exactly their requirements.

Q5. Are the Avatars frameworks delivering virtual frameworks in order to provide e-learning effective?

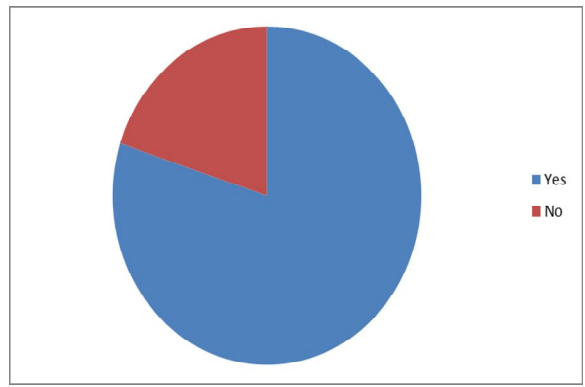

Figure 5. Responding with usage of Avatar

Discussion: If considering the outcome achieved with the responses received through the survey conducted by the respondents, it is clear that Avatars are effective. 80% of the respondents are with the Avatars frameworks however, following the web 2.0 technology for having online activities to be carried out but are as good as other social networking sites. The E-learning process essentially requires a strong system like an Avatars framework and such other technology that is updated with the current new requirements of learners and educators worldwide.

## 4. Findings

The findings of the research study made over the effectiveness of e-learning process are made clear with the responses given by the respondents. The responses to the questionnaire filled by the 150 respondents randomly selected from a randomly chosen university of such environment. Findings of the research finally received are recorded with the help of statistical tools such as bar charts, diagrams, pie charts etc. These charts are the best ways to depict the respondents in a true sense that could be analysed easily by the experts. Inference provided from the responses provided below at each question is good to give an overall idea for the entire study.

## 5. Discussion

Discussions based on the findings of the research study made over the topic of effectiveness of e-learning highly the concerns of the people. The questions given in the questionnaires for responding over the qualities and concern of the instructors and professors for the learners are point of discussions. Questions in the questionnaire are directly prepared on the basis of the respondents opinion views about anything that is achieved with technology used in the e-learning process.

The asynchronous nature of the e-learning process should be made more effective by the use of the latest technology. MATEP tools should be enhanced with more user friendly technology that satisfies the needs of both instructors and learners throughout the world.

These tools have to be equipped with technology that could provide clearer interactions, virtual efficiency and better simulation than at present. E-learning is necessary for future generations that would use this technology in the far flung areas of the world. Modern Teaching aids in the online framework should be able to provide e-learning support through an e-tutoring and e-mentoring system. In this way timely individualised learning support and feedback can be provided to the student anytime, anywhere and any place; so as to best support the achievement of learning objectives and enhanced learning quality of experience.

## 6. Conclusion

E-learning is the step forward to secure future educational needs with modern teaching aids efficiently. As a conclusion of the entire presented study undertaken as part of this work, it may be said that virtual environment interaction such as e-learning and monitoring with avatar is beneficial for the program. A comparative analysis of different platforms to provide an optimised delivery of content has been presented. Avatars and similar applications are the most advanced technology of current times with the ability to deliver and monitor student activities. Although MATEP has drawbacks to be improved such as the need for integrated data mining models to build significant student patterns by themselves, nonetheless, it poses to be the best monitoring and analysis tool provider that is used by e-learners in the virtual world by helping the instructor to acquire a more accurate view of what is happening in distance learning. From the conducted study and subsequent analysis of the impact of LMS integrated learning path for activity tracking and performance evaluation, this paper concludes that model-driven personalisation support modalities are beneficial in providing enhanced levels of learning and trusted assimilation in an e-learning delivery context.

## REFERENCES

[1] P. Henry, "E-learning technology, content and services", Vol. 43. No. 4. pp. 249-255, 2001.
[2] G. Attwell, "Evaluating E-learning: A Guide to the Evaluation of E-learning, Evaluate Europe Handbook Series", Vol. 2, pp.5-41, 2006.
[3] C. Quinn, "Designing mLearning: Tapping into the Mobile Revolution for Organizational Performance", San Francisco: Pfeiffer, USA, 2011.
[4] M.A. Sicilia, "Competencies in Organizational E-Learning: Concepts and Tools", University of Alcala, pp.352-366, 2007.
[5] P. Scott, C. Vanoirbeek, "Technology-Enhanced Learning", Technology-Enhanced Learning, Vol. 71(1), pp. 12-13, 2007.
[6] S. Naidu, C. Vanoirbeek, "E-Learning: Definition, Scope, Trends, Attributes & Opportunities", A Guidebook of Principles, Procedures and Practices, Commonwealth of Learning, pp. 12-32, .. 2006.
[7] S. Chou, C. Liu, "Learning effectiveness in a web-based virtual learning environment: A learner control perspective", Journal of Computer Assisted Learning, Journal of Computer ...... Assisted Learning, 2005.
[8] M. May, S. George, P, Prevot," TrAVis to Enhance Students' Self-monitoring in Online lLearning Supported by Computer-Mediated Communication Tools", International Journal of Computer Information Systems and Industrial Management Application, Vol. 3 pp. 623-634, 2011.
[9] T. Mayes, S.D. Freitas "JISC e-Learning Models Desk Study, Stage 2: Review of e-learning theories", frameworks and models, 2006.
[10] S. Britain, "A Review of Learning Design: Concept, Specifications and Tools A report for the JISC E-learning Pedagogy Programme", Vol. pp. 3-21, 2004.

International Journal of Computer Science & Information Technology (IJCSIT) Vol 6, No 3, June 2014

[11] E. Popescu, D.Cioiu, "Instructors Support for Monitoring and Visualizing Students' Activity in a Social Learning Environment", 12th IEEE International Conference on Advanced Learning Technologies, pp. 1-3, 2012.
[12] Burdescu, Mihaescu"Building Intelligent E learning systems by Activity Monitoring and Analysis", available at: http://link.springer.com/chapter/10.1007%2F978-3-642-13396-1_7#page-2, 2006.
[13] M.E. Zorrilla, E. Álvarez, "MATEP: Monitoring and Analysis Tool for e-Learning Platforms", Eighth IEEE International Conference on Advanced Learning Technologies, pp. 1-3, 2008.
[14] J. Martino, "The avatar project: connected but not engaged—the paradox of cyberspace", available at: http://art.tafe.vu.edu.au/avatar/wp-content/uploads/AvatarLitReview- revision%202.doc, 2007.
[15] S. Woolgar, "Virtual Society? Technology, Cyberbole, Reality", Oxford: Oxford University Press,2002.
[16] J.E. Katz, R.E. RiceHenry, "Social consequences of internet use: access, involvement and interaction", MIT Press, 2002.
[17] A.J.P. van den Brekel, "Get your consumer health information from an avatar!: health and medical related activities in a virtual environment", available at:http://www.bm.cm.uj.krakow.pl/eahil/proceedings/oral/vanBrekel.pdf, Retrieved June 6, 2008.
[18] G. Falloon, "Using avatars and virtual environments in learning: What do they have to offer?", British Journal of Educational Technology, Vol. 41, No, pp.108–122., 2010.
[19] S. Yucel, "E-Learning Approach in Teacher Training", E-Learning Approach in Teacher Training, Vol. 7 No.4, pp. 123-129, 2006.
[20] Mazzoni & Gaffuri, "Monitoring Activity in E-Learning: A Quantitative Model Based on Web Tracking and Social Network Analysis", Universtiy of Bolonga, pp. 10-20, 2010.
[21] B. Jong,T. Chan, Y. Wu, "Learning Log Explorer in E-Learning Diagnosis", IEEE Transactions on education, Volume: 50, No.3, pp. 216-228, 2007.
[22] M. May, S. George, P. Prévôt, "A Closer Look at Tracking Human & Computer Interactions in Web-Based Communications", International Journal of Interactive Technology and Smart Education, Vol.5, No.3, pp. 170-188, 2008.

## AUTHORS

Ali Alowayr is a Ph.D student in Reading University, at the School of Systems Engineering. Alowayr is the head of Computer Science department in Albaha University, Saudi Arabia. He lecturered at the same department and taught several courses: Advanced Data base ( SQL), C++ Programming Language, Web Internet applications, Human Computer Interaction. Alowayr worked as a supervisor on the Balanced Scorecard program, to implement the Strategic Plan of King Saud University, Saudi Arabia in 2011.

Atta Badii is a professor at the University of Reading where he is the Director of the Intelligent Systems Research Laboratory, at the School of Systems Engineering. Atta is Director of the European Virtual Centre of Excellence for Ethically-guided and Privacy-respecting Video-Analytics (VideoSense), Director of the European Observatory for Crowd-Sourcing and Collective-Awareness Platforms for Transformative Government (Citizens' Say), Chair of the International Companion Robotics Institute (CRI), and, the Internet of People Things and Services (IoPT) Research Forum as well as the European Privacy by Co-Design Research Cluster.